# Thermoelectric properties of in-plane $90^0$-bent graphene nanoribbons with nanopores

**Van-Truong Tran[1] and Alessandro Cresti[2]**

[1]IMPMC, Université Pierre et Marie Curie (UPMC), Sorbonne Universités, 75252 Paris Cedex 05, France

[2]Univ. Grenoble Alpes, Univ. Savoie Mont Blanc, CNRS, Grenoble INP, IMEP-LAHC, 38000 Grenoble, France

E-mail: vantruongtran.nanophys@gmail.com and alessandro.cresti@grenoble-inp.fr

**Abstract**

We study the thermoelectric performance of $90^0$-bent graphene nanoribbons containing nanopores for optimized design of multiple functional circuits including thermoelectric generators. We show that the thermal conductance of the $90^0$-bent ribbons is lower from few times to an order of magnitude compared to that of pristine armchair and zigzag straight ribbons. Consequently, the thermoelectric performance of the bent ribbons is better than its straight ribbon counterparts, in particular at high temperatures above 500 K. More importantly, the introduction of nanopores is demonstrated to strongly enhance their thermoelectric capacity. At 500 K, the figure of merit $ZT$ increases by more than 160% (from 0.39 without pores to 0.64) with 3 nanopores incorporated, and by more than 200% (up to 0.88) when 24 nanopores are introduced. $ZT \approx 1$ can be achieved at a temperature of about 1000 K. In addition, the thermoelectric performance is shown to be further improved by adopting asymmetrical leads. This study demonstrates that $90^0$-bent ribbons with nanopores have decent thermoelectric performance for a wide range of temperatures and may find application as efficient thermoelectric converters.

## 1. Introduction

Electronics has seen a huge advance over the past half-century thanks to the massive and rapid progress in the downscale size of transistors, which helps to design more compact electronic circuits and therefore smaller devices with better performance [1]. However, heat management in small electronic devices has been facing many challenges due to important self-heating effects, which can shorten the lifetime of the device [2,3]. Together with the search for better thermal conductors and efficient designs to dissipate heat in devices, the possibility to harvest the wasted heat and convert it into electric energy represents an important challenge [4,5], which has inspired the design of circuits with multiple functions including thermoelectric generators.

Nanostructures based on graphene and other 2D materials have been considered to be potential for compact and efficient electronic components in circuits thanks to their atomic thickness and outstanding physical properties [3,6–11]. In particular, although the natural thermoelectric ability of graphene is poor [12], this material is still versatile for thermoelectric applications thanks to its high melting point (estimated to be >4000 K [13]), and as its nanostructuration allows the engineering of its electronic and thermal properties, thus providing highly tunable thermoelectric performance [12,14–17]. In particular, graphene nanoribbons (GNRs) show better thermoelectric performance compared to 2D sheets thanks to the presence of a band gap, which leads to a larger Seebeck coefficient [12]. By engineering GNRs with isotopes doping [16], nanopores [18–20], vacancies [21,22], or edge roughness [23–25], the figure of merit *ZT* of GNRs can be tuned in a wide range from 0.1 to 4.

The best thermoelectric performance of graphene-based systems observed so far is in nanostructured straight ribbons [16,20,23,26]. However, to integrate different systems into a circuit for multiple functions, the arrangement of different components might require the use of bent ribbons to optimize the space in the circuit [27,28]. Due to the strong variation of both electronic and thermal properties of graphene ribbons with specific edge orientation, the thermoelectric performance of bent ribbons could vary significantly compared to their straight counterparts. Although a few works already studied the electronic [29–31], thermal [32,33], and thermoelectric properties of in-plane bent ribbons [34], their level of efficiency in thermoelectric conversion still needs investigation.

In this work, we focus on the improvement of the thermoelectric performance of ribbons bent with an angle of $90^0$, which exhibits the lowest thermal conductance compared to other angles [32]. To further reduce the thermal conductance and optimize the thermoelectric performance, we introduce nanopores. We also consider asymmetrical leads to maximize the mismatch between in-coming and out-going phonon modes for additional degradation of the phonon conductance. Our goal is to optimize $90^0$-bent ribbons to make them suitable as efficient thermoelectric components in multiple functional circuits.

The rest of the paper is organized as follows. In Sec. 2, we describe the system under investigation, introduce the tight-binding (TB) model for electrons and the Force Constant (FC) model for phonons, and present the Non-Equilibrium Green's Functions (NEGF) formalism for the simulation of transport properties. Section 3 illustrates and discusses our numerical results. In Sec. 3.1, we focus on the thermoelectric ability of $90^0$-bent ribbons containing a single nanopore. In Sec. 3.2, we investigate the variation of the electronic, thermal and thermoelectric properties of bent ribbons in the presence of multiple nanopores. In Sec. 3.3, the impact of asymmetrical leads is discussed. Finally, conclusions are given in Sec. 4.

## 2. Investigated device and methodology

### 2.1. Investigated device

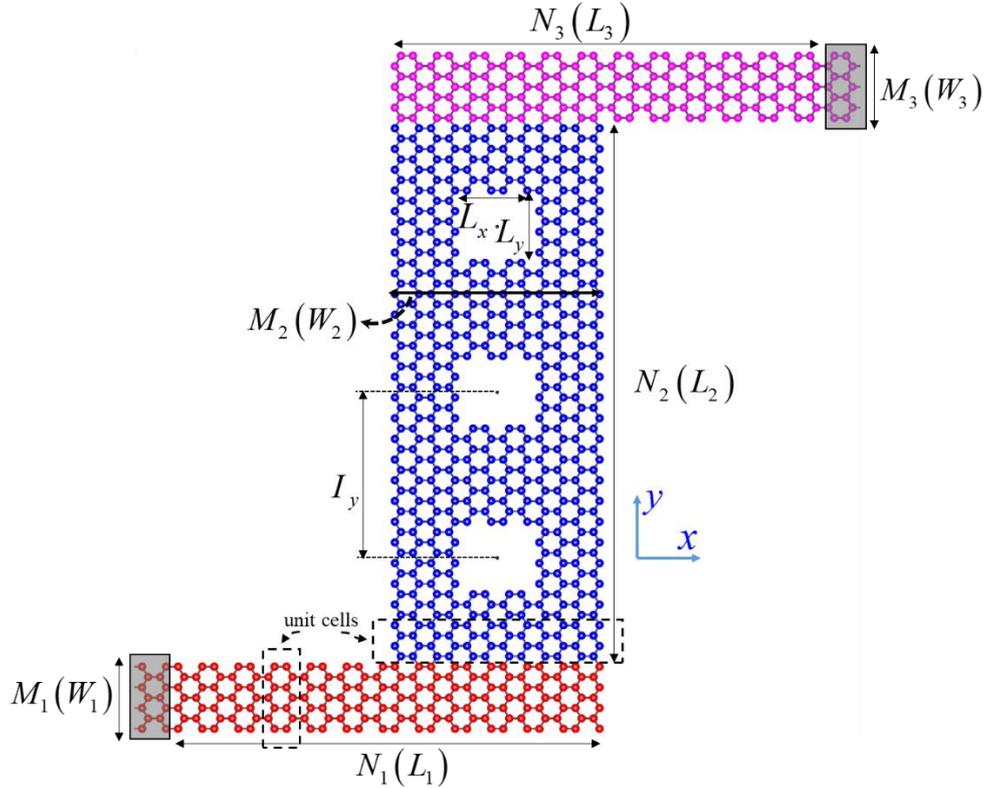

**Figure 1.** Sketch of a $90^0$-bent graphene nanoribbon composed of two horizontal armchair-edge sections and a vertical zigzag-edge section. Size of the ribbon parts: $M_1 = 7$ ($W_1 \approx 0.74$ nm), $N_1 = 12$ ($L_1 \approx 5.11$ nm); $M_2 = 12$ ($W_2 \approx 2.41$ nm), $N_2 = 13$ ($L_2 \approx 6.39$ nm); $M_3 = 7$ ($W_3 \approx 0.74$ nm), $N_3 = 12$ ($L_3 \approx 5.11$ nm). Size of nanopores: $L_x = L_y = 5\ a_0$, $I_y = 13.86\ a_0$, $n_{pores} = 3$. The boxes with dashed lines indicate the unit cells of the horizontal and vertical ribbons, which contain 2 and 4 slices *per* cell, respectively.

The geometry of a typical device based on a $90^0$-bent graphene nanoribbon is illustrated in figure 1, where the two horizontal segments are armchair graphene ribbons (AGNRs) and the vertical part is a zigzag graphene ribbon (ZGNR). For each section, the size is defined by the width of the ribbon and its length along the ribbon axis. The width of an AGNR section is $W_A = (M-1) \times \sqrt{3}/2 \times a_0$, and that of a ZGNR section is $W_Z = (3M-2) \times a_0/2$, where $a_0 = 0.142$ nm is the distance between two nearest carbon atoms and $M$ is the number of dimer (zigzag) lines across the ribbon width of an AGNR (ZGNR) [35]. The bottom (top) horizontal ribbon contains $N_1$ ($N_3$) cells, each composed of 2 slices perpendicular to the $x$-axis (see the box with dashed lines) and with a length $L_1 = 3 \times a_0 \times N_1$ ($L_3 = 3 \times a_0 \times N_3$). The two horizontal sections directly connect to the leads (marked as gray boxes). The thermal current flows from the hot contact to the cold contact, while the electric current flows in the opposite direction. It is worth noting the horizontal parts decouple the leads from the scattering region in the simulation, and the results do not depend on $L_1$ and $L_3$ when these lengths are long enough. The vertical ribbon is along the $y$-axis and consists of $N_2$ cells each composed of 4 slices (in the box with dashed lines) for a total length $L_2 = 2\sqrt{3} \times a_0 \times N_2$. The rectangular nanopores introduced to reduce the phonon conductance have size $L_x$x$L_y$ and spacing $I_y$. The nanopores are located in the middle of the vertical ribbon. In the structure sketched in figure 1, we have $M_1 = M_3 = 7$ ($W_1 = W_3 \approx 0.74$ nm), $N_1 = N_3 = 12$ ($L_1 = L_3 \approx 5.11$ nm), $M_2 = 12$ ($W_2 \approx 2.41$ nm), $N_2 = 13$ ($L_2 \approx 6.39$ nm), and number of nanopores $n_{pores} = 3$, each with $L_x = L_y = 5\, a_0$, and distance $I_y = 13.86\, a_0$. Ribbons of such a width can be successfully synthesized by growth in ultrahigh vacuum [36], and nanopores can be created by high energy electron beam [37,38] or by direct chemical growth [39].

## 2.2. Methodology

To investigate the thermoelectric properties of bent ribbons, we employed a TB model to describe electrons, an FC model to describe phonons, and the NEGF formalism for simulating the transport properties of both electrons and phonons.

### Model for electrons

In the present work, we adopted a TB model with coupling up to the third nearest neighbors (3NN) to describe electrons. This model properly describes the electron-hole asymmetry of the electronic structure. The TB Hamiltonian reads:

$$\begin{cases} H_e = \sum_i \varepsilon_i |i\rangle\langle i| - \sum_{i,j \in \overline{1,3\text{NN}}} t_{ij} |i\rangle\langle j| \\ s_{ij} = \langle i | j \rangle \end{cases} \quad (1)$$

where the state $|i\rangle$ corresponds to the $p_z$ orbital of the $i$-th carbon atom, $\varepsilon_i$ is the on-site energy, and $t_{ij}$ and $s_{ij}$ are the inter-atomic hopping and overlap parameters, respectively. The TB parameters were optimized for graphene ribbon structures based on the fitting of *ab-initio*-calculated electronic structures [35].

*Model for phonons*

To study phonon properties, a four-nearest-neighbor-coupling FC model was employed, which accurately reproduces the phonon dispersion of graphene obtained by *ab initio* methods and experimental measurements [40]. The secular equation for phonons resulting from Newton's second law reads [41,42]:

$$DU = \omega^2 U, \quad (2)$$

where $U$ is the matrix containing the vibrational amplitudes of all atoms, $\omega$ is the angular frequency, and $D$ is the dynamical matrix [41]:

$$D = \left[ D_{3\times 3}^{ij} \right] = \left[ \begin{cases} -\dfrac{K_{ij}}{\sqrt{M_i M_j}} & \text{for } j \neq i \\ \sum_{j \neq i} \dfrac{K_{ij}}{M_i} & \text{for } j = i \end{cases} \right], \quad (3)$$

with $M_i$ the mass of the $i$-th atom, and $K_{ij}$ the tensor coupling between the $i$-th and $j$-th atoms defined by the unitary rotation in the plane:

$$K_{ij} = U^{-1}(\theta_{ij}) K^0_{ij} U(\theta_{ij}), \quad (4)$$

where $U(\theta_{ij})$ is the rotation matrix [42], $\theta_{ij}$ is the angle between the positive direction of the $x$-axis and the vector from the $i$-th atom to the $j$-th atom, and $K^0_{ij}$ is the force constant tensor given by:

$$K^0_{ij} = \begin{pmatrix} \Phi_r & 0 & 0 \\ 0 & \Phi_{t_i} & 0 \\ 0 & 0 & \Phi_{t_o} \end{pmatrix}, \quad (5)$$

with $\Phi_r, \Phi_{t_i}, \Phi_{t_o}$ the coupling components in radial, transverse in-plane and out-plane, respectively, which are given in terms of the 12 parameters of the FC model, whose values were taken from Wirtz's work [40]. The mass of carbon atoms is $1.994\times10^{-26}$ kg.

### *Simulations of the transport properties*

To investigate the transport properties of both electrons and phonons, we employed the NEGF technique [43]. All structures were considered as composed of three parts: the left and right leads and the device (central) region. The leads were considered as semi-infinite periodic regions, with the same unit cell as that of the two horizontal ribbons. The device region contains the two horizontal ribbons and the vertical part.

The retarded Green's function for electrons can be written as:

$$G = \left[ E^+ S_D - H_D - \Sigma^s_L - \Sigma^s_R \right]^{-1}, \quad (6)$$

where $E^+ = E + i\eta$ with $E$ the energy and $\eta$ is a positive infinitesimal number, and

$$\Sigma^s_L = \left( E^+ S_{DL} - H_{DL} \right) G^0_L \left( E^+ S_{LD} - H_{LD} \right)$$
$$\Sigma^s_R = \left( E^+ S_{DR} - H_{DR} \right) G^0_R \left( E^+ S_{RD} - H_{RD} \right) \quad (7)$$

are the surface self-energies contributed from the left and right contacts. $G^0_{L(R)}$ is the surface Green's function of the isolated left (right) contact. To calculate the self-energy, a modified Sancho-Rubio iterative technique to include the overlap factor was adopted [44,45].

For the phonon calculation, we employed a similar method obtained just by replacing the energy $E$ by $\omega^2$, and $H_D$, $H_{DL}$, $H_{LD}$, $H_{DR}$, $H_{RD}$ by $D_D$, $D_{DL}$, $D_{LD}$, $D_{DR}$, $D_{RD}$, respectively. We also considered that $S_D = \mathbf{1}, S_{DL} = S_{LD} = S_{DR} = S_{RD} = \mathbf{0}$ for phonons.

The recursive technique of ref. [46] was employed to efficiently manage the size of the device in the Green's function calculation. Then electron (phonon) transmission was computed as [41,43]:

$$T_{e(p)} = Trace\left\{ \Gamma^s_L \left[ i\left( G_{11} - G_{11}^\dagger \right) - G_{11} \Gamma^s_L G_{11}^\dagger \right] \right\}, \quad (8)$$

where $\Gamma^s_{L(R)} = i\left( \Sigma^s_{L(R)} - \Sigma^{s\,\dagger}_{L(R)} \right)$ is the surface injection rate at the left (right) contact. The electrical conductance $G_e$, the Seebeck coefficient $S$, the electron thermal conductance $K_e$ at given chemical potential $\mu$ were computed with the Landauer-Onsager's approach [47] as

$$G_e(\mu,T) = e^2 L_0(\mu,T)$$

$$S(\mu,T) = \frac{1}{e}\frac{1}{T}\frac{L_1(\mu,T)}{L_0(\mu,T)} \quad (9)$$

$$K_e(\mu,T) = \frac{1}{T}\left[L_2(\mu,T) - \frac{L_1(\mu,T)^2}{L_0(\mu,T)}\right]$$

where the intermediate functions $L_n$ are defined as [41]

$$L_n(\mu,T) = \frac{1}{h}\int_{-\infty}^{+\infty} dE\, T_e(E)(2K_b T)^{n-1} g^e_n(E,\mu,T), \quad (10)$$

with $g^e_n(E,\mu,T) \equiv \left(\frac{E-\mu}{2k_B T}\right)^n / \cosh^2\left(\frac{E-\mu}{2k_B T}\right)$ and $k_B$ the Boltzmann constant.

The phonon thermal conductance $K_p$ was computed by [41]

$$K_p = \frac{K_b}{2\pi}\int_0^{\infty} d\omega\, \mathrm{T}_p(\omega) g^p(\omega,T), \quad (11)$$

where $g^p(\omega,T) \equiv \left(\frac{\hbar\omega}{2k_B T}\right)^2 / \sinh^2\left(\frac{\hbar\omega}{2k_B T}\right)$. Since the phonon thermal conductance is mainly limited by the extremely reduced size of the bent ribbon, the bending itself, and the presence of nanopores, we can reasonably disregard the contribution from anharmonic phonon-phonon scattering, which is not included in our model.

The figure of merit *ZT* was calculated from the electrical conductance, the Seebeck coefficient, the electron and phonon thermal conductances as [12,41,48]

$$ZT = \frac{G_e S^2}{K_e + K_p}T. \quad (12)$$

## 3. Results and discussions

In this section, first the electronic, thermal and thermoelectric properties of bent ribbons without and with a single nanopore are compared with those of straight ribbons. Then, additional nanopores are introduced in the system to further enhance phonon scattering. Finally, the role of asymmetrical leads on the thermoelectric performance of bent ribbons is discussed.

### 3.1. $90^0$-bent ribbons with a single nanopore

We examine the thermal, electronic and thermoelectric properties of bent ribbons without and with a single nanopore.

Figure 2(a) shows that the phonon thermal conductance of the bent-ribbon structures (red and blue lines) is about an order of magnitude smaller than that of the straight ZGNR (black line), and a few times smaller than that of the straight AGNR (violet line). At 500 K, the conductance of the considered straight zigzag ($M = 12$), armchair ($M = 7$) and corresponding $90^0$-bent ribbons without and with a square pore with $L_x = L_y = 5\ a_0$ is 5.89, 1.63, 0.71, 0.53 nW/K, respectively.

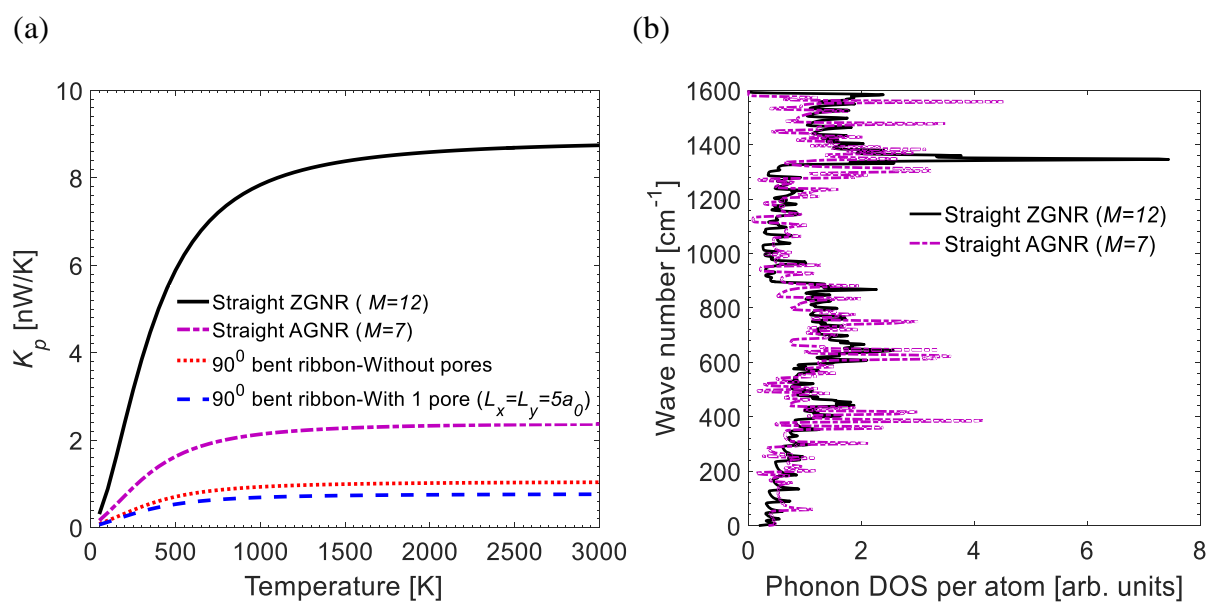

**Figure 2**. (a) Phonon thermal conductance as a function of temperature of the bent ribbon without (red) and with (blue) a nanopore in comparison with those of the straight ZGNR (solid black) and AGNR (dashed-dotted violet). Size of the bent ribbons: $M_1 = M_3 = 7$ ($W_1 = W_3 \approx 0.74$ nm); $M_2 = 12$ ($W_2 \approx 2.41$ nm), $N_2 = 7$ ($L_2 \approx 3.44$ nm). Size of the nanopore: $L_x = L_y = 5\ a_0$, $n_{pores} = 1$. (b) Phonon DOS (in arbitrary unit) as a function of the wave number for the periodic straight AGNR and ZGNR.

The reduction of the phonon conductance of the bent ribbons can be explained by the mismatch between phonon modes of the armchair and zigzag sections, as seen by a comparison of the phonon density of states (DOS) for periodic straight armchair and zigzag ribbons. The phonon DOS was calculated by using the Gaussian smearing of the Dirac delta function [49], i.e., $DOS(\omega) = \sum_{n} \sum_{\vec{k} \in BZ} \delta(\omega - \omega_n(\vec{k})) \approx \sum_{n} \sum_{\vec{k} \in BZ} \frac{1}{\eta\sqrt{\pi}} e^{-\frac{(\omega - \omega_n(\vec{k}))^2}{\eta^2}}$, where $n$ is the phonon band index, $\vec{k}$ is a wave vector in the first Brillouin zone (BZ), $\eta$ is a small positive number, and $\omega_n(\vec{k})$ is the frequency of the $n$-th phonon mode at the wave vector $\vec{k}$, which is calculated from equation (2). As can be seen from the different positions of the peaks (i.e., of the van Hove singularities) in figure 2(b), the discrepancy between the phonon states of the armchair and zigzag ribbons is remarkable, in particular, in the range from 0 to 900 cm$^{-1}$ and from 1300 cm$^{-1}$ to 1600 cm$^{-1}$.

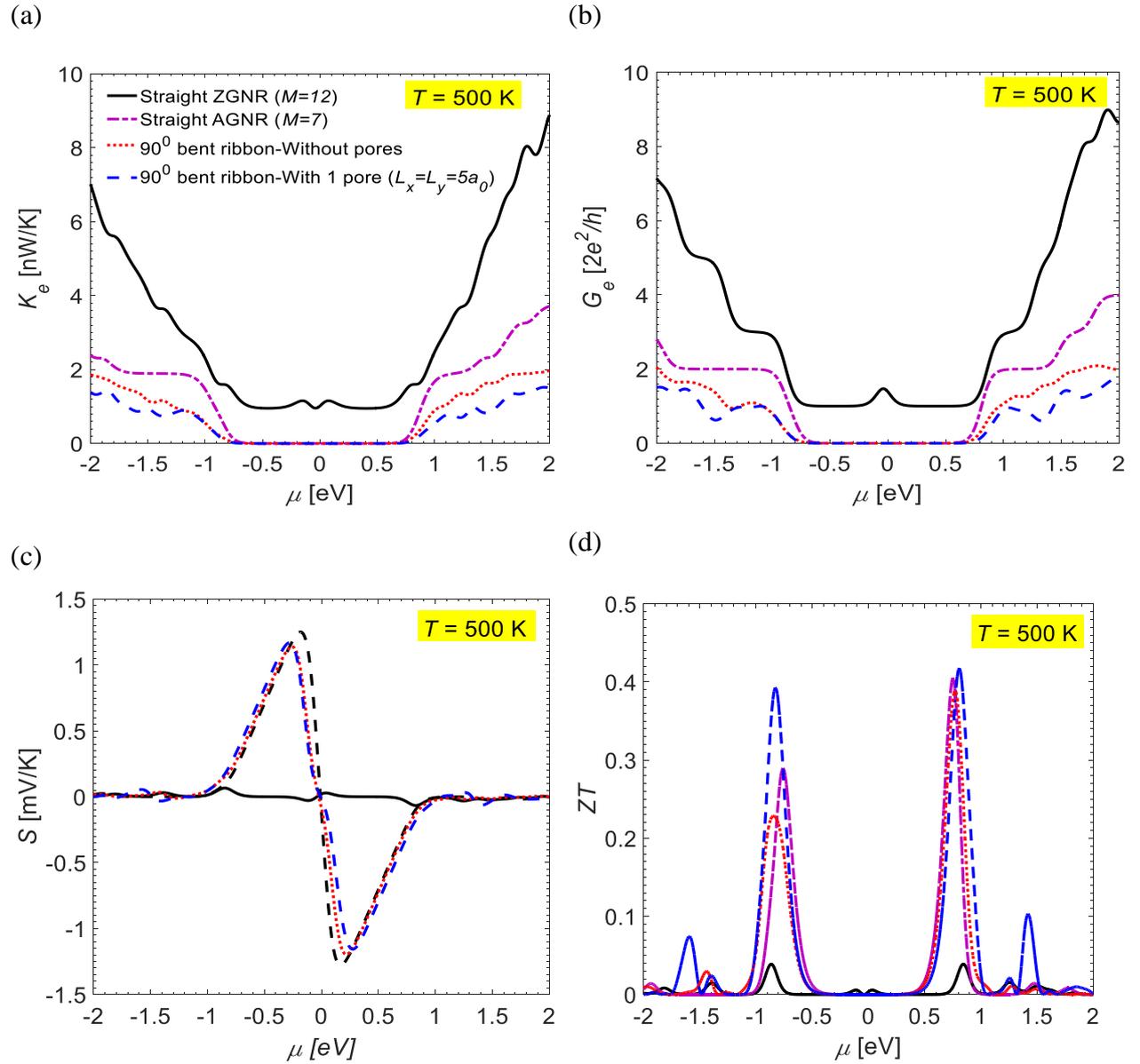

**Figure 3**. (a) Electron thermal conductance, (b) electrical conductance, (c) Seebeck coefficient, and (d) *ZT* at 500 K as a function of chemical potential for the bent ribbon without (red) and with (blue) a nanopore in comparison with those of the straight ZGNR (solid black) and AGNR (dashed-dotted violet). Size of the bent ribbons: $M_1 = M_3 = 7$ ($W_1 = W_3 \approx$ 0.74 nm); $M_2 = 12$ ($W_2 \approx 2.41$ nm), $N_2 = 7$ ($L_2 \approx 3.44$ nm). Size of the nanopore: $L_x = L_y =$ 5 $a_0$, $n_{\text{pores}} = 1$.

Figures 3(a), 3(b) and 3(c) present the electron thermal and electrical conductance and Seebeck coefficient as a function of the chemical potential $\mu$ at 500 K for the four studied systems. We chose a reference temperature of 500 K because, as illustrated later, starting from this value a significant figure of merit $ZT \geq 0.4$ can be achieved. Due to the semi-metallic behavior, indicated by the non-zero electronic conductance over the whole range of energy, the straight zigzag structure has a very small Seebeck coefficient (see the black solid line in figure 3(c)) below 66 $\mu$V/K. In contrast, the armchair structure $M = 7$ and the bent-ribbon counterparts all present a decent gap in the electric conductance of about 1.3 eV. Despite the change in the direction along the ribbon axis, which causes scattering (for both electrons and phonons) and then a degradation of the electric conductance, the transport gap in these bent structures is observed to be similar to that of the straight AGNR. This indicates that the global transport gap of the bent-ribbon system is determined by the device section with the largest band gap, which corresponds to the AGNRs. As a result, the Seebeck coefficient of the bent structures is similar to that of the corresponding straight AGNRs, as shown in figure 3(c).

The figure of merit *ZT* at 500 K as a function of chemical potential, see figure 3(d), reveals that, due to the small Seebeck coefficient, the straight ZGNR has a *ZT* smaller than 0.04. The maximum $ZT_{\text{max}}$ (among those corresponding to different chemical potentials) of the straight AGNR and of the bent ribbon without and with a square nanopore is 0.41, 0.39 and 0.42, respectively, and can be obtained for $\mu < 1$ eV.

At high temperatures, $ZT_{\text{max}}$ is significantly higher for bent ribbons than for straight ribbons, as seen in figure 4(a). For example, at 1400 K, the straight ZGNR has $ZT_{\text{max}} \approx 0.09$, while this value is about 0.77, 0.96 and 1.03 for the straight AGNR and the bent ribbons without and with the pore, respectively. The effect is even more pronounced at higher temperatures. It is worth mentioning that, below 2000 K, the bent ribbon with the nanopore always exhibits better thermoelectric performance than that without the pore. This is a consequence of the lower thermal conductance, as observed in figure 2(a). The decrease of $ZT_{max}$ above 2200 K can be understood from the fact that the electron thermal conductance $K_e$, considered at the chemical potential $\mu$ corresponding to $ZT_{max}$, increases with increasing temperature (figure 4(b)), while the phonon thermal conductance $K_p$ tends to saturate at high temperatures, see figure 2(a). Therefore, above a certain temperature, $K_e$ dominates over $K_p$ as can be seen in the inset of figure 4(b) for the case of the bent ribbon without pores. Similar results were observed for the other structures. Once $K_e$ is comparable with $K_p$, and it even becomes the leading term contributing to the total thermal conductance, $ZT_{max}$ decreases, as seen in all the curves in figure 4(a).

It is also instructive to analyze the temperature dependence of the electrical conductance and the Seebeck coefficient at the chemical potential corresponding to $ZT_{max}$. The results shown in figures 4(c) and 4(d) indicate that, except for the electrical conductance of the ZGRN structure, which increases remarkably with increasing temperature, the electrical conductance and the absolute Seebeck coefficient of the other structures fluctuate modestly with temperature. As the phonon and electron thermal conductances overall rise with temperature as shown in figures 2(a) and 4(b), these results reveal that the increase of $ZT_{max}$ with increasing temperature as seen in figure 4(a) is indeed due to the increase of temperature factor $T$ that appears in the $ZT$ definition, see equation (12). Moreover, the electrical conductance of the bent ribbons is observed to be smaller than that of the straight ribbons, and the absolute Seebeck coefficient of these bent structures is close to that of the straight armchair ribbon at most temperatures. Therefore, the lower phonon thermal conductance observed in figure 2(a) is the major advantage of the bent ribbons, which leads to higher $ZT$ values compared to those of straight ribbons.

It is also worth noting that the negative sign of the Seebeck coefficients in figure 4(d) implies that $ZT_{max}$ occurs for positive chemical potentials (see figure 3(c)) at most temperatures, therefore these structures can be classified into the *n*-class thermoelectric materials. Interestingly, from 1000 K to 2600 K, the bent ribbon with a nanopore can change to the *p*-class as revealed by the positive sign of the Seebeck coefficient observed in that temperature range.

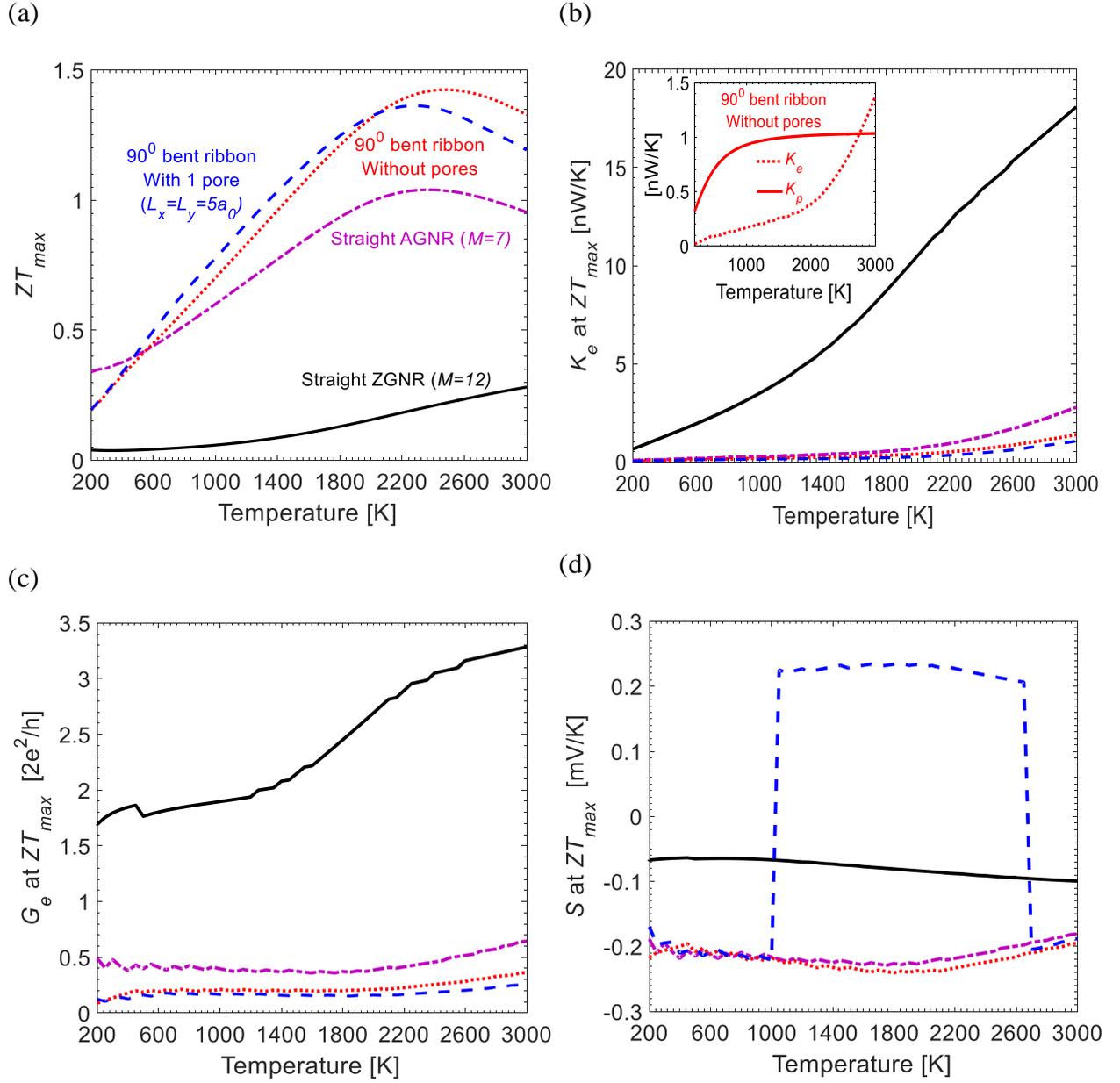

**Figure 4**. (a) $ZT_{\max}$, (b) electron thermal conductance $K_e$, (c) electrical conductance $G_e$, (d) Seebeck coefficient $S$ at the chemical potential corresponding to $ZT_{\max}$, as a function of temperature for the bent ribbon without (red) and with (blue) a nanopore in comparison with those of the straight ZGNR (solid black) and AGNR (dashed-dotted violet).

Based on the previous literature about the impact of edge disorder on the electronic and thermal properties of straight graphene ribbons, we expect that, even in the presence of disorder along the bent-ribbon edges, whose structure may also be deformed at the high temperatures we consider, it is still possible to achieve high thermoelectric properties. Indeed, an edge disordered ribbon is likely to show a transport gap either due to the natural bandgap of its armchair-like segments, or due to the scattering stemming from the mismatch between its armchair and zigzag sections [50,51] or due to Anderson localization in the area of defects [52]. Even in the case of edge reconstruction, a gap is still expected in the conductance profile of the armchair ribbon sections [53]. Most importantly, regarding the phonon properties, Sevinçli *et al.* [23] demonstrated that the thermal conductance decreases strongly in the presence of edge disorder, which results in an enhanced thermoelectric performance. A quantitative study of the impact of edge disorder on the *ZT* factor of our system is out of the scope of the present study.

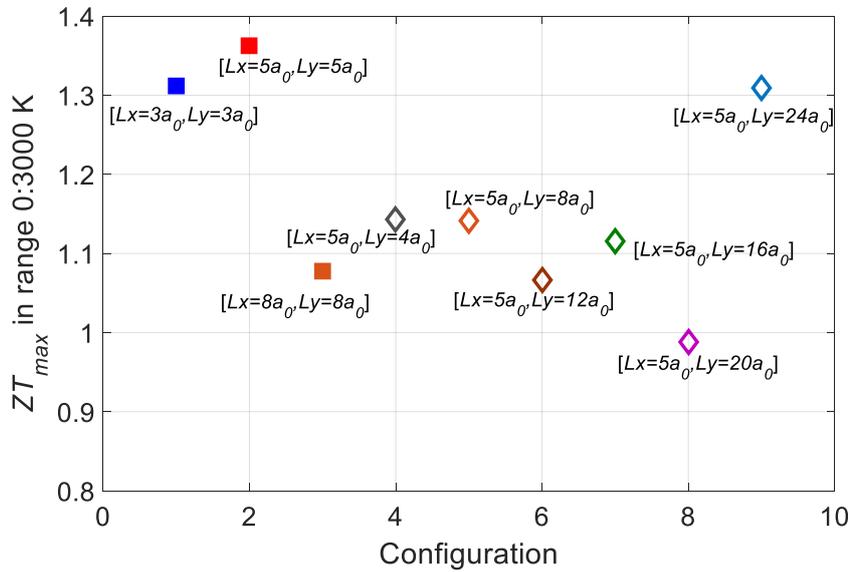

**Figure 5**. $ZT_{max}$ in a wide range of temperature from 0 to 3000 K for different nanopore sizes. Size of the bent ribbons: $M_1 = M_3 = 7$ ($W_1 = W_3 \approx 0.74$ nm); $M_2 = 12$ ($W_2 \approx 2.41$ nm), $N_2 = 7$ ($L_2 \approx 3.44$ nm). Only one nanopore is present.

We also explored the dependence of the thermoelectric ability of the bent ribbon on the nanopore shape and size. Figure 5 shows that, among the considered configurations, the bent ribbon with the square nanopore of size $L_x = L_y = 5\ a_0$ provides the best $ZT_{max} \approx 1.36$ (where $ZT_{max}$ corresponds to the highest *ZT* value calculated for different chemical potentials and over a temperature range between 0 and 3000 K). It is worth noting that also other configurations have $ZT_{max}$ larger than 1. As also commented in the next section, by reducing the width of the active region in the vertical section of the device and acting as a scatterer, a large nanopore significantly affects the phonon thermal conductance, but also degrades the electrical conductance. The identified optimal size for the nanopore is the one that, for the given geometry, provides the best trade-off between these two aspects.

### *3.2. $90^0$-bent ribbons with multiple nanopores*

In this section, the effect of multiple nanopores in the vertical part of the bent ribbons is examined as a function of their size and distance $I_y$.

In figure 6(a), the phonon conductance of the bent ribbons with 3 nanopores is shown. Different square pore sizes and three values of the $I_y$ parameter are considered. The phonon thermal conductance strongly varies with the nanopore size, in agreement with the previous studies with nanopores in straight ribbons [15,54]. Such size effects stem from the fact that the pore size along the width of the vertical ribbon is directly associated with the number of phonon modes and with the cross section transferring heat in the region with nanopores. Therefore, it has a remarkable impact on the thermal conduction, which, in general, is smaller for larger nanopores. In contrast, the phonon thermal conductance does not vary significantly with the distance $I_y$. The slight difference observed between configurations with different $I_y$ could originate from the phonon coherence in ring-like structures, where constructive or destructive phonon interference could occur [55]. Such interference is even more pronounced in the case of electrons and is commonly referred to as quantum interference [56]. The $ZT_{max}$ of these structures is shown in figure 6(b) as a function of temperature. Although the structures with the largest pore considered here have a lower phonon conductance, the $ZT_{max}$ is found higher in a wide range of temperature for the intermediate nanopore size $L_x = L_y = 5$ $a_0$, as already observed for the case of a single nanopore.

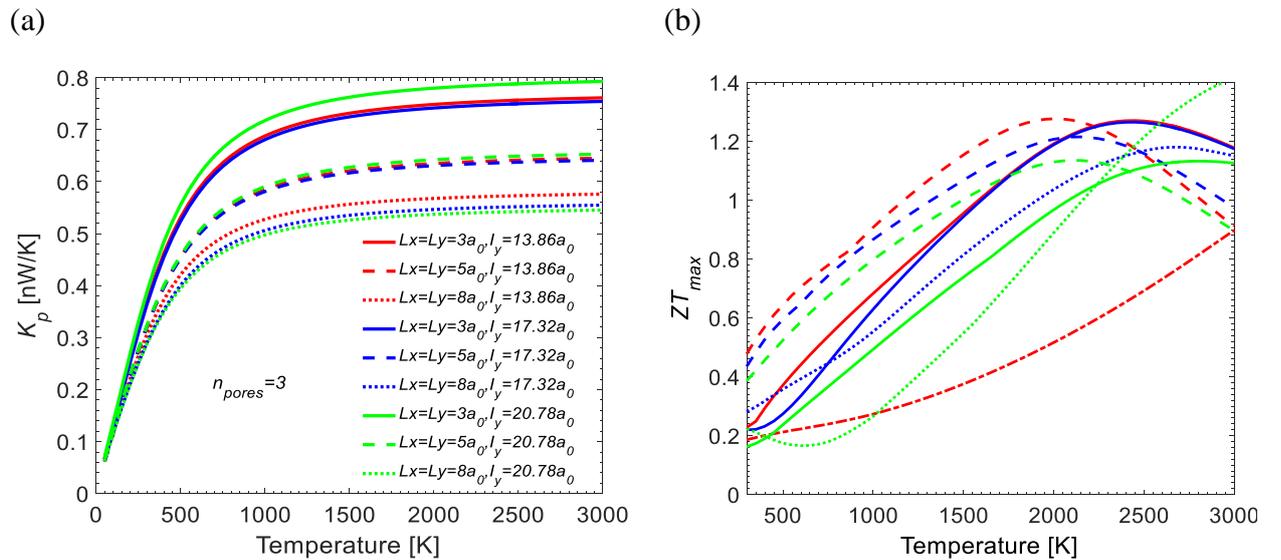

(a)          (b)

**Figure 6.** (a) Phonon thermal conductance and (b) $ZT_{max}$ as a function of temperature for different square nanopores at distance $I_y$. The studied structures contain 3 nanopores. Size of the bent ribbons: $M_1 = M_3 = 7$ ($W_1 = W_3 \approx 0.74$ nm); $M_2 = 12$ ($W_2 \approx 2.41$ nm), $N_2 = 13$-$17$ depending on $I_y$.

To see how the thermoelectric performance could be impacted by varying the number of nanopores $n_{pores}$ in the vertical part, we considered different lengths of the vertical ZGNR section. Figure 7(a) shows the corresponding $ZT_{max}$ as a function of temperature. The thermoelectric performance for temperature below 1000 K is overall enhanced when increasing $n_{pores}$. A figure of merit $ZT$ of about 1.4 can be achieved with the structure containing $n_{pores} = 12$ at about 2000 K. In figure 7(b), $ZT_{max}$ is plotted as a function of the number of pores with the optimal size $L_x = L_y = 5\ a_0$ at 300 K (diamonds), 500 K (squares) and 1000 K (triangular). The results indicate that $ZT_{max}$ below 1000 K is enhanced significantly from $n_{pores} = 0$ to $n_{pores} = 3$, then it increases gradually and eventually tends to saturate. This can be understood as, for $n_{pores} > 3$, the vertical part tends to behave as a superlattice of nanopores [15,41]. At 300 K, $ZT_{max}$ can be enhanced about 1.85 times from 0.26 without pores to 0.48 with $n_{pores} = 3$, and $ZT_{max}$ reaches 0.69 with $n_{pores} = 24$. $ZT_{max}$ at 500 K also increases remarkably from 0.39 without pores to 0.64, 0.71, 0.76 and 0.88 with $n_{pores} = 3, 6, 12$, and 24, respectively.

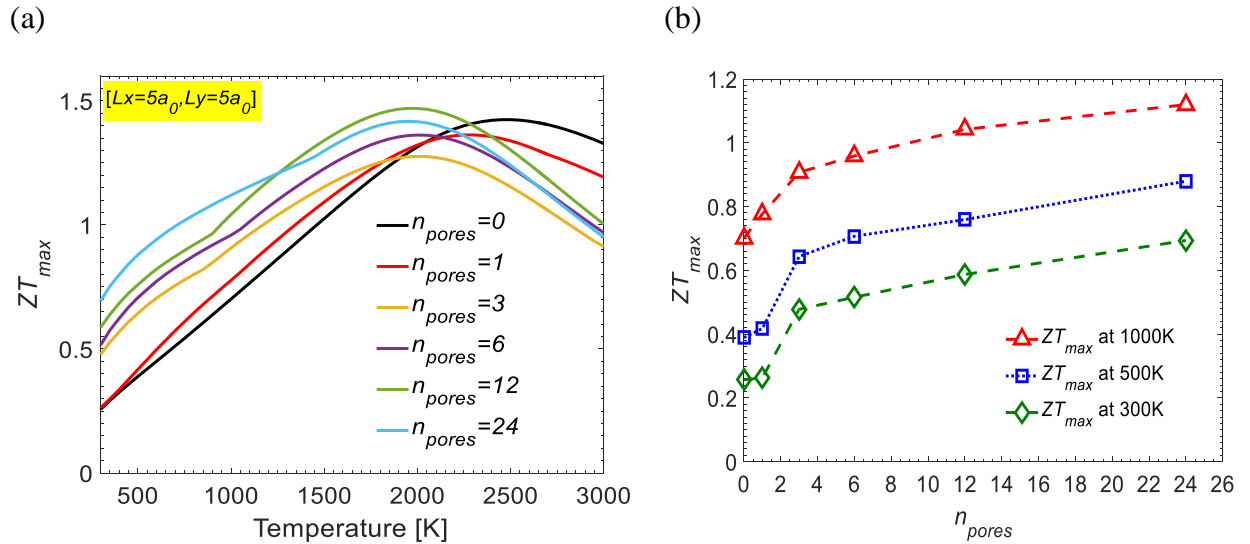

**Figure 7**. (a) $ZT_{max}$ as a function of temperature with a different number of nanopores incorporated in the systems. (b) $ZT_{max}$ as a function of the number of nanopores at different temperatures. The pore size is fixed for all the structures $L_x = L_y = 5\ a_0$. The size of the bent

ribbons: $M_1 = M_3 = 7$ ($W_1 = W_3 \approx 0.74$ nm), $M_2 = 12$ ($W_2 \approx 2.41$ nm), $N_2$ ($L_2$) varies with the number of nanopores.

It is worth noting that the electronic properties of AGNRs are classified into 3 groups with the number of dimer lines across the ribbon equal to $3p$, $3p+1$, $3p+2$ with $p$ in an integer number [57]. As seen in figure 3(b), the transport gap of the bent ribbons without and with nanopores is directly related to the transport gap of the armchair sections. Therefore, it is interesting to investigate how the thermoelectric capacity of bent ribbons depends on the AGNR groups.

Figure 8(a) reports $ZT$ at 500 K as a function of chemical potential for three ribbons of width $M_1 = M_3 = 7$, 8 and 9, which represent the $3p+1$, $3p+2$ and $3p$ groups, respectively. $ZT$ clearly shows the highest peak with the structure of width $M = 7$, due to the fact that the $3p+1$ group provides the largest band gap [57]. Interestingly, although the $3p+2$ group features the smallest gap among the three groups, the thermoelectric performance of the system with $M = 8$ turns out to be better than for the system with $M = 9$ (group $3p$). As in the case of the nanopore size, this indicates the importance of the trade-off between the phonon thermal conductance and the Seebeck coefficient around the Fermi level, which is found to be better for the structure with horizontal AGNR sections belonging to the $3p+2$ group.

Figure 8(b) reports $ZT_{max}$ in a wide range of temperatures from 0 to 3000 K for different structures of these three AGNR groups. The result confirms that the $3p+1$ group provides higher thermoelectric performance compared to the other groups.

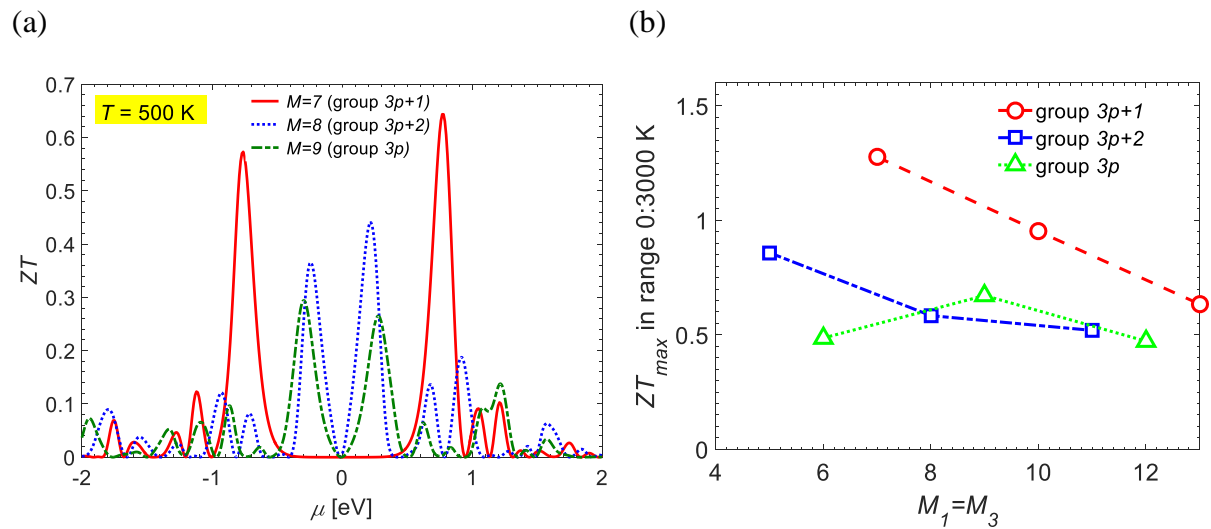

**Figure 8**. (a) $ZT$ at 500 K as a function of chemical potential for bent ribbons with different width $M = M_1 = M_3$ of the horizontal AGNR segments. (b) $ZT_{max}$ in a wide range of temperature from 0 to 3000 K for several structures of the three AGNR groups with $M = 3p$, $3p+1$, and $3p+2$. Other parameters: $M_2 = 12$ ($W_2 \approx 2.41$ nm), $N_2 = 13$ ($L_2 \approx 6.39$ nm), $L_x = L_y = 5\ a_0$, $I_y = 13.86\ a_0$, $n_{pores} = 3$.

## *3.3. Asymmetrical leads*

The structures considered up to now have symmetrical leads, i.e., the same width of the lower and upper horizontal ribbons. It might also be worth investigating structures with asymmetrical leads, as this can induce further phonon mode mismatch for in-coming and out-going phonon modes in the two leads, and thus a possible higher thermoelectric performance due to a lower thermal conductance.

Figure 9(a) shows $ZT_{max}$ as a function of temperature for several structures with different pairs of widths [$M_1$, $M_3$] and 3 nanopores. The results reveal that the correlation between the sizes of the two leads strongly impacts the thermoelectric performance of the systems. Among the considered structures, that with the two leads of width $M_1 = 7$ and $M_3 = 11$ exhibits the highest figure of merit $ZT$, in particular, compared to the symmetrical leads $M_1 = M_3 = 7$ and $M_1 = M_3 = 11$. The discrepancy in the results of the two configurations [$M_1$, $M_3$] equal to [7,11] and [11,7] origins from the asymmetry of the vertical ZGNR about the *x*-axis (see figure 1), which leads to a difference in couplings between the vertical section and the lower and upper horizontal ribbons.

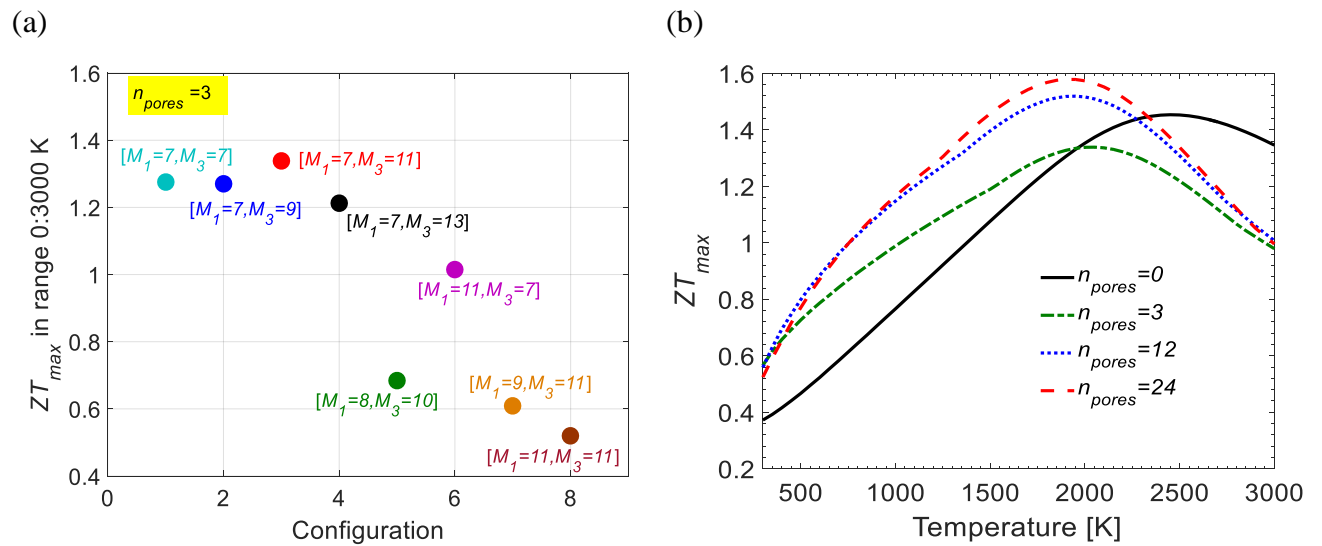

**Figure 9**. Bent ribbons with asymmetrical lead: (a) $ZT_{max}$ in a wide range of temperatures from 0 to 3000 K is considered for different configurations of asymmetrical leads and with 3 nanopores. (b) $ZT_{max}$ in structures with $M_1 = 7$ and $M_3 = 11$ as a function of temperature for different numbers of nanopores incorporated in the systems. In all structures, the nanopores have size $L_x = L_y = 5\ a_0$, $I_y = 13.86\ a_0$ and the width of the vertical ribbon is fixed at $M_2 = 12$ ($W_2 \approx 2.41$ nm).

Further examination for the asymmetrical lead structures with $M_1 = 7$, $M_3 = 11$, $ZT_{max}$ as a function of temperature for different numbers of incorporated nanopores is shown in figure 9(b). A $ZT_{max} \approx 1.6$ could be reached with asymmetrical lead systems with $n_{pores} = 24$, which is larger than that of the corresponding system with symmetrical leads, as seen in figure 7(a). Compared to the asymmetrical-lead system without nanopores (solid black line), the introduction of nanopores significantly increases $ZT_{max}$ for a wide range of temperature. For example, at 500 K, $ZT_{max}$ increases from 0.46 with $n_{pores} = 0$ to 0.73 and 0.78 for $n_{pores} = 3$ and 24, respectively.

It is noticed that we did not consider here the role of graphene-electrode interfaces, which could affect both the thermal and the electrical conductance. While the $ZT$ factor would benefit from a degradation of the former, it might be hindered by a reduction of the latter. The impact is expected to be negligible for contacts placed over large graphene regions seamlessly and monolithically joined to the armchair ribbons [27] (away from the bent ribbon), for which the resulting contact resistance [58] is expected to be much smaller than that of the bent ribbon. On the contrary, contacts directly made over the narrow ribbons could entail a resistance that is larger than that of our system [59] and whose impact evaluation would require further investigations, which are out of the scope of the present study.

## 4. Conclusion

The introduction of nanopores in $90^0$-bent graphene nanoribbons allowed us to predict a significant improvement of their thermoelectric properties, which are determined by the subtle interplay between electric and phonon transport.

For pristine systems without pores, we demonstrated that the phonon thermal conductance of the bent ribbon was about an order of magnitude smaller than that of straight zigzag ribbons, and a few times smaller than that of straight armchair ribbons. On the other side, the electrical conductance showed a general decrease and a transport gap close to the band gap of the armchair horizontal sections. Due to such electronic and thermal characteristics, the thermoelectric of the bent ribbon resulted remarkably higher than that of its straight zigzag and armchair counterparts, in particular above 500 K.

By introducing nanopores in the bent structures, the thermoelectric performance was further enhanced thanks to the reduction of the phonon thermal conductance. At 500 K, the *ZT* of the bent ribbon increased from 0.39 without nanopores to up to 0.88 in the presence of 24 nanopores. *ZT* values above 1 could be achieved starting from a temperature of about 1000 K. The role of the width and symmetry of the leads, which consisted of armchair ribbons, was also explored. In particular, armchair ribbons belonging to the 3*p*+1 group turned out to show the highest *ZT*. Finally, the thermoelectric performance was shown to be further improved by combining asymmetrical leads, which decreased the phonon thermal conductance by introducing additional phonon mode mismatch.

Our results provide an insight into the thermoelectric capacity of $90^0$-bent graphene ribbons and pave the way to enhance their thermoelectric performance by including nanopores in the structure. Our findings with structures based on graphene can be also extended to other 2D materials such as silicene, a two-dimensional material composed of Si atoms [60]. Silicene has an electronic structure [60] that is similar to that of graphene, but a few times lower thermal conductance (due to a narrower range of phonon frequencies [61]). Therefore, bent silicene ribbons with nanopores, even with an additional introduction of isotope disorder [62], could lead to a significantly high figure of merit *ZT*.